\documentclass[conference]{IEEEtran}
\IEEEoverridecommandlockouts
\usepackage{etoolbox,xstring,mfirstuc,textcase}
\usepackage{cite}
\usepackage{amsmath,amssymb,amsfonts}
\usepackage{algorithmic}
\usepackage{graphicx}
\usepackage{textcomp}
\usepackage[hidelinks]{hyperref} 
\usepackage{xcolor}
\usepackage{algorithm}
\usepackage{multirow}
\usepackage{makecell}
\usepackage{booktabs}
\usepackage{amsmath}
\usepackage[caption=false,font=normalsize,labelfont=sf,textfont=sf]{subfig}
\usepackage{tabularx}
\usepackage{diagbox}
\usepackage{mdframed} 
\definecolor{mycolor}{RGB}{64,105,224}
\usepackage{pifont}
\def\BibTeX{{\rm B\kern-.05em{\sc i\kern-.025em b}\kern-.08em
		T\kern-.1667em\lower.7ex\hbox{E}\kern-.125emX}}
\begin{document}

\title{ModFus-DM: Explore the Representation in Modulated Signal Diffusion Generated Models\\
		
\thanks{This research was supported in part by the National Natural Science Foundation of China under Grant 62401429 and 62425113, in part by the China Postdoctoral Science Foundation under Grant Number GZC20241332 and 2025M771739, in part by the Seed Fund for Technology Clusters of the 10th Research Institute, China Electronics Technology Group Corporation, 2024JSQ0203 . \textit{Corresponding author: Zhenxi Zhang.} }
\thanks{The authors are with the Xidian University, Xian 710126, China (e-mail: 21021210621@stu.xidian.edu.cn; yli\_1999@stu.xidian.edu.cn; zhangzhenxi@\\xidian.edu.cn; xrshi@xidian.edu.cn;	 fzhou@mail.xidian.edu.cn.)}
}

	\author{Haoyue Tan, Yu Li, Zhenxi Zhang*, Xiaoran Shi, Feng Zhou}
	
	\maketitle
	
\begin{abstract}
Automatic modulation classification (AMC) is essential for wireless communication systems in both military and civilian applications. However, existing deep learning-based AMC methods often require large labeled signals and struggle with non-fixed signal lengths, distribution shifts, and limited labeled signals. To address these challenges, we propose a modulation-driven feature fusion via diffusion model (ModFus-DM), a novel unsupervised AMC framework that leverages the generative capacity of diffusion models for robust modulation representation learning. We design a modulated signal diffusion generation model (MSDGM) to implicitly capture structural and semantic information through a progressive denoising process. Additionally, we propose the diffusion-aware feature fusion (DAFFus) module, which adaptively aggregates multi-scale diffusion features to enhance discriminative representation. Extensive experiments on RML2016.10A, RML2016.10B, RML2018.01A and RML2022 datasets demonstrate that ModFus-DM significantly outperforms existing methods in various challenging scenarios, such as limited-label settings, distribution shifts, variable-length signal recognition and channel fading scenarios. Notably, ModFus-DM achieves over 88.27\% accuracy in 24-type recognition tasks at SNR$\ge$12dB with only 10 labeled signals per type.
\end{abstract}
	
	\begin{IEEEkeywords}
	Automatic modulation classification, diffusion feature, diffusion generation model, self-supervised learning.
	\end{IEEEkeywords}

\section{Introduction}
Automatic modulation classification (AMC) aims to autonomously identify the modulation type of radio signals in unknown or non-cooperative environments. It serves as a critical prerequisite for subsequent processes such as demodulation, interference suppression, and signal analysis \cite{10224342_Zhang}. AMC techniques are widely applied in domains such as cognitive radio \cite{4600222_Wu}, electronic reconnaissance \cite{10752883}, and spectrum management \cite{9395503_Bhatti}. 
Traditional AMC methods often depend on pre-designed and handcrafted expert features \cite{tan_pass-net_2024}, which limits their effectiveness in modern wireless environments characterized by complexity, variability, and high labeling costs. 


Deep learning (DL)-based AMC approaches have gained widespread adoption due to their remarkable performance, implementation simplicity, and deployment efficiency \cite{10296032_Qian, 10461997, 11054047}. 
Despite their success on benchmark datasets, most existing DL-based AMC methods rely heavily on large-scale labeled datasets with fixed structures to train end-to-end recognition models. However, real-world wireless communication systems involve complex environments and diverse signal sources, where obtaining abundant, high-quality labeled signals is both costly and impractical \cite{tan_multi-scale_2024, 10949625_Wang}, significantly limiting model performance. Moreover, most existing DL-based models assume fixed-length signals, making them poorly suited to the variable-length sequences commonly encountered in practical scenarios. Consequently, leveraging unlabeled signals for modulation representation learning (MRL) and designing flexible architectures have become key research priorities.


\begin{figure}[!t]
	\centering
	\captionsetup{skip=0pt}
	\setlength{\abovecaptionskip}{0cm}
	\setlength{\abovecaptionskip}{0cm}
	\includegraphics[width=0.5\textwidth]{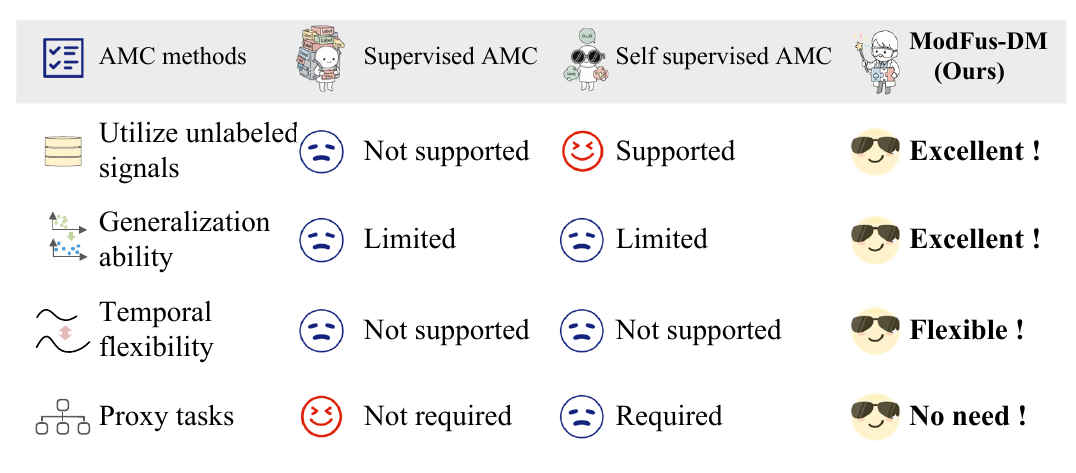}
	\vspace{-0.5cm}
	\caption{Comparison of ModFus-DM with traditional supervised and self-supervised AMC methods across four key aspects.}
	\label{Ability}
	\vspace{-0.5cm}
\end{figure}

Self-supervised learning (SSL) methods learn representations by constructing proxy tasks from unlabeled data. Common approaches to proxy task design include data augmentation-based strategies \cite{kong_transformer-based_2023, xiao_mclhn_2024} and transformation-based contrastive learning (CL) \cite{li_unsupervised_2023}. However, the performance of these SSL methods is inherently constrained, as the learned representations are highly dependent on the design of suitable proxy tasks—which can vary significantly across different data distributions. Moreover, these methods often fail to comprehensively model the signal evolution process and underlying distributional characteristics. In non-cooperative and dynamically changing communication environments, where signal sources are abundant and prior knowledge is scarce, designing effective proxy tasks becomes particularly challenging. As a result, the generalizability and robustness of SSL-based AMC methods remain substantially limited.

Can models move beyond reliance on proxy tasks and directly learn robust representations from raw, unlabeled signals? A promising direction is to leverage tasks that require a comprehensive understanding of the data itself. Among these, generative tasks are particularly compelling, as they inherently demand the model to capture both the global structure and fine-grained details of the input. This motivates the exploration of generative models under unsupervised learning paradigms. In contrast to task-specific discriminative models, generative models focus on capturing the intrinsic structural patterns of data without requiring explicit labels. If a model can successfully reconstruct or generate data, it implies that it has effectively internalized the essential semantic and structural features, which inherently encode valuable discriminative information.

Traditional generative models, such as variational autoencoders (VAEs) \cite{VAE_2014} and generative adversarial networks (GANs) \cite{goodfellow_generative_2014}, have achieved notable success in generation tasks but also exhibit inherent limitations. The latent space of VAEs is typically constrained by a prior Gaussian distributions. GANs, on the other hand, suffer from unstable training, and their discriminators primarily focus on distinguishing real from fake samples rather than learning generalizable structural features. As a result, both VAEs and GANs offer limited structural expressiveness and fail to capture multi-scale, temporal, and fine-grained patterns. 

Diffusion models, with their iterative denoising paradigm, offer a promising pathway for MRL without relying on labels or proxy tasks. Unlike traditional generative models that generate signals in an end-to-end manner, diffusion models adopt a progressive refinement approach. This allows models effectively capture multi-scale temporal features and suppress non-structural interference, which is well aligned with the complexity and variability of real-world electromagnetic environments \cite{chen_data_2024, li_diffusion_2025}. 
From another perspective, this process resembles an archaeological excavation: layer by layer, noise and irrelevant patterns are stripped away, gradually revealing core structural features essential for modulation recognition.  
These features often emerge and become amplified at initial stages, enhancing the model’s discriminative power and generalization ability. Moreover, their temporal flexibility and contextual awareness make them well-suited for recognizing variable-length signals and handling distribution shifts.

While diffusion models offer promising representational capabilities, effectively extracting representative features from them remains a key challenge for downstream recognition tasks. In diffusion models, the forward process progressively corrupts the signal, while the reverse process typically employs a U-Net noise prediction network, to iteratively denoise and reconstruct signals \cite{li_diffusion_2025, song_denoising_2022}. Intermediate layers within the U-Net capture semantically and structurally meaningful representations, making them natural candidates for feature extraction.
Prior studies have shown that both the encoder and decoder paths in U-Net contain rich discriminative information \cite{xiang2023denoising, yang2023diffusion}. However, due to variations in data distribution, signal complexity, and intrinsic modulation patterns, the model’s reliance on specific feature layers may vary across tasks. Consequently, identifying an optimal extraction layer remains a challenging problem.
To address this, we propose a diffusion-aware feature fusion (DAFFus) module that adaptively aggregates hierarchical and multi-scale features across U-Net layers. By dynamically emphasizing the most informative representations, DAFFus addresses the challenge of optimal layer selection and establishes a more robust, task-adaptive foundation for downstream recognition. 


In this paper, we propose a novel unsupervised MRL framework, named modulation-driven feature fusion via diffusion model (ModFus-DM), which effectively exploits latent discriminative features from unlabeled signals. Specifically, we propose modulated signal diffusion generation models (MSDGM) to facilitate representation learning by progressively reconstructing unlabeled signals from noise. This process enables the model to capture modulation patterns and implicitly learn distributional characteristics. Furthermore, we design the DAFFus mechanism, which adaptively integrates multi-scale and multi-stage intermediate features from MSDGM, enhancing the extraction of discriminative representations. To highlight the advantages of our approach, we provide a comparative capability analysis in Fig. \ref{Ability}.

The key contributions of our paper are as follows:
\begin{itemize}
\item We propose a novel unsupervised modulation representation learning framework, ModFus-DM, which eliminates the reliance on large amounts of labeled data and proxy tasks. This approach is well-suited for complex, dynamic, and non-cooperative communication environments where annotation is scarce.
\item We develop the MSDGM, which progressively reconstructs signals through iterative denoising. This process enables the model to implicitly capture the structural and semantic characteristics of modulation patterns, thereby laying a robust foundation for modulation representation learning.
\item We design the DAFFus mechanism, which adaptively integrates intermediate features across multiple scales and stages within the reverse process. This enhances the quality and robustness of the learned representations by emphasizing task-relevant information. 
\item Extensive experiments on four benchmark datasets demonstrate that ModFus-DM consistently outperforms existing methods in limited-label recognition, distribution shifts scenarios, variable-length signal tasks and real-world scenarios, highlighting its effectiveness and generalizability.
\end{itemize}

\vspace{-0.3cm}
	\section{Related Work}
	\subsection{Traditional DL-AMC}
	In recent years, deep learning has demonstrated superior performance in AMC \cite{10857965, 10261289}. Existing DL-based AMC methods can be broadly categorized into three primary architectures. First, convolutional neural network (CNN)-based structures \cite{oshea_over--air_2018, liu_deep_2017}, extract spatial features from modulated signals to achieve classification. Due to their simple architecture and efficiency, these methods were widely adopted in early AMC research. Second, recurrent neural network (RNN)-based architectures \cite{ke_real-time_2022} focus on the temporal characteristics of signal sequences, capturing modulation features by modeling time dependencies. Third, hybrid structures that integrate CNNs and RNNs \cite{xu_spatiotemporal_2020} jointly capture spatial and temporal information to establish a more robust feature space. In addition to these, complex-valued networks \cite{liu_automatic_2023} and self-attention-based architectures \cite{kong_transformer-based_2021, kong_transformer-based_2023} also show promising performance. The former directly model the amplitude and phase components of complex signals, improving modulation feature representation. The latter dynamically focus on critical segments of the temporal sequence via attention mechanisms.
	
	However, these data-driven, fully supervised models face two significant challenges. First, the complexity of the real-world communication environment and the diversity of signal sources make acquiring large-scale, and accurately labeled signals exceedingly costly, accurately labeled datasets prohibitively expensive, greatly limiting their practical applicability. Second, most models are designed under the assumption of fixed-length inputs, lacking adaptability to variable-length signal sequences encountered in real-world communication systems, thereby reducing their robustness. 

	\subsection{Self-supervised AMC}
	
	SSL enhances feature extraction by designing proxy tasks that enable models to learn representations from unlabeled data. Recent advances on self-supervised AMC have primarily centered around contrastive learning frameworks, with various proxy tasks formulated to optimize feature space. D. Liu et al. \cite{liu_self-contrastive_2021} constructed positive and negative sample pairs using signal rotation operations, leveraging rotational invariance to capture modulation characteristics. W. Kong et al. \cite{kong_transformer-based_2023} proposed a transformer-based contrastive learning framework that employs time-warping augmentation to improve feature robustness. Y. Fu et al. \cite{fu_hybrid-view_2025} introduced contrastive learning across multiple perspectives, including raw signal domain, spectrogram transformed via wavelet analysis, and ``star video" representation, aiming to maximize inter-domain consistency. Additionally, Y. Li et al. \cite{li2025multi} integrated intra-domain and inter-domain information, where the former was generated through various signal augmentation techniques, and the latter derived from different transformations. 
	
	However, existing approaches fall within the scope of proxy task-based SSL methods, which inherently depends on carefully designed augmentations and transformations. This reliance poses challenges for modulation recognition in non-cooperative scenarios, motivating further exploration of generative-based SSL methods for AMC.

\subsection{Diffusion Models for AMC}
\begin{figure*}[!t]
	\centering
	\captionsetup{skip=0pt}
	\setlength{\abovecaptionskip}{0cm}
	\setlength{\abovecaptionskip}{0cm}
	\includegraphics[width=0.95\textwidth]{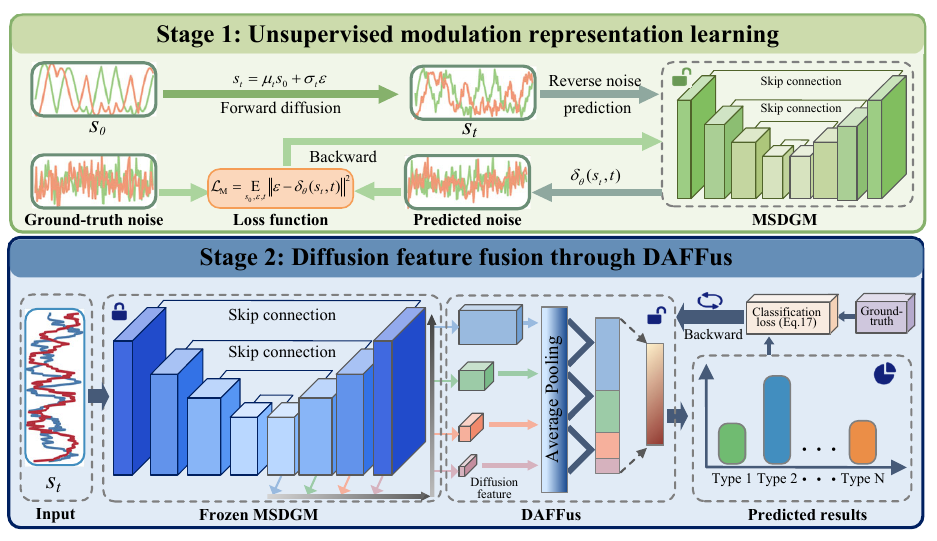}
	\caption{Overall structure of ModFus-DM framework. During stage 1, MSGDM capture the signal structure and modulation semantics using unlabeled signals through both forward diffusion and reverse processes. In stage 2, the DAFFus module is optimized using a limited amount of labeled data, while the MSDGM remains frozen. This two-stage process enables effective effective MRL and achieves accurate recognition without the need of pre-designed proxy tasks or fine-tuning of the generative model.}
	\label{OverallStructure}
	\vspace{-0.5cm}
\end{figure*}
Although diffusion models have outstanding generative performance in visual and perceptual domains, such as image \cite{song_denoising_2022} and video \cite{ho2022video} synthesis, their application in wireless signal processing remains in an exploratory stage. 
Due to their high-quality generative capabilities, diffusion models have recently been applied to modulation recognition tasks, primarily as data augmentation techniques. For instance, J. Chen et al. \cite{chen_data_2024} developed a diffusion model constrained by class labels and used the generated signals to augment the original training dataset. Similarly, M. Li et al. \cite{li_diffusion_2025} utilized multi-step intermediate samples from the reverse diffusion process as augmented data. These methods effectively expand training datasets and improve the performance of recognition models.

However, the existing AMC methods utilize diffusion models as generative tools, often overlooking the rich and robust information embedded within the diffusion process. In contrast, our study adopts a novel perspective by investigating the representational capacity and discriminative potential of diffusion models for modulation recognition. Our work opens new research directions for applying diffusion models in the AMC domain.
	
\section{Problem Definition}
The objective of the AMC task is to identify the modulation type of a received signal, serving as a fundamental prerequisite for subsequent processes such as demodulation and signal analysis. AMC plays a particularly crucial role in communication systems, especially under non-cooperative conditions where parameters such as carrier frequency and channel state are unknown. The received modulated signal can be represented as:
\begin{equation}
s[n] = \alpha  \cdot r[n - \tau ] \cdot {e^{j(2\pi \Delta fn{T_s} + \phi )}} + w[n]
\label{eq_sig_moddel}
\end{equation}
where $r[n]=\mathcal{M}\left( x[n] \right)$ denotes the modulated baseband signal, where $\mathcal{M}$ is the modulation function, and $ x[n]$ is the original information sequence. $\alpha $ denotes the channel fading coefficient, $\tau$ is the timing offset, $\Delta f$ represents the carrier frequency offset, $T_s$ indicates the sampling interval, $\phi$ is the initial phase offset, and $w[n]$ represents additive Gaussian noise.
	
Self-supervised modulation feature extraction eliminates the need for labeled data by learning representations of different modulation types directly from unlabeled signals. Let the unlabeled dataset be denoted as ${\mathcal{D}}_u = \{ s_i[n] \}_{i=1}^N$, where $i$ is the signal index and $N$ is the total number of signals. Through an SSL strategy, a well-trained representation function ${f_{\theta^*}}( \cdot )$ is obtained to map signal into a latent feature space:
\begin{equation}
{z_i} = {f_{\theta^* }}({s_i}[n])
\end{equation}
where $\theta *$ denotes the parameters of the trained model. Signals of the same modulation type cluster closely in the representation space, while those of different types are more distinctly separated. Guided by a classifier, the representation ${z_i}$ is then mapped into a probability space to complete the AMC task.

\section{The Proposed Method}
\subsection{Overall Structure}

The overall structure of the proposed ModFus-DM framework is illustrated in Fig. \ref{OverallStructure}, which comprises two main stages: self-supervised MRL and diffusion feature fusion via DAFFus. In the self-supervised MRL stage, the MSDGM operates directly on raw modulated signals. Through a forward diffusion process and a reverse noise prediction process, MSDGM implicitly learns the underlying structure and modulation patterns of the signals. This iterative denoising process enables the model to learn rich, hierarchical semantic representations of modulation without relying on any label information. Building on the learned diffusion feature space, the DAFFus module adaptively integrates features across different semantic levels within the diffusion model. By selectively emphasizing important diffusion features, DAFFus constructs a robust and discriminative feature space tailored to the recognition task. Finally, the modulation type is predicted.

\subsection{Modulated Signal Diffusion Generated Models}
\begin{figure}[!t]
	\centering
	\captionsetup{skip=0pt}
	\setlength{\abovecaptionskip}{0cm}
	\setlength{\abovecaptionskip}{0cm}
	\includegraphics[width=0.5\textwidth]{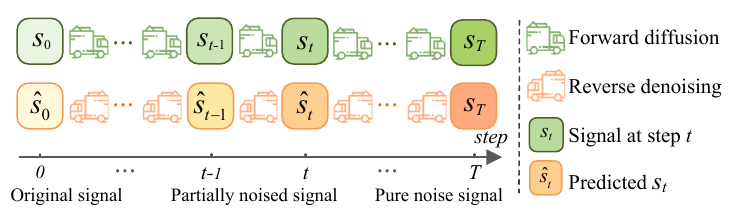}
	\caption{Forward diffusion and backward denoising process. 
	By learning to reconstruct $s_0$ from $s_T$, the model captures the structural characteristics and semantic representations of the modulation patterns embedded in the signal.}
	
	\label{DiffusionProcess}
	\vspace{-0.5cm}
\end{figure}
Unlike conventional self-supervised modulation representation learning approaches that rely on meticulously designed proxy tasks, the proposed MSDGM is trained solely based on the intrinsic structure of modulated signals. It follows a bidirectional process, where raw signals are progressively diffused into noise during the forward process and subsequently reconstructed through reverse denoising. On one hand, learning directly from the signal mitigates reliance on predefined proxy tasks, enhancing adaptability across varying signal scenarios. On the other hand, the iterative denoising implicitly captures hierarchical structural and semantic features of the signals, establishing a solid foundation for recognition tasks.

The self-supervised diffusion feature learning process comprises a forward diffusion stage and a reverse noise prediction stage. The forward process serves as a continuous noise injection mechanism applied to the original signal and and is formulated as a Markov process. At each diffusion step, a small amount of noise is introduced into the original signal $s_0$, progressively degrading its inherent structural information. After \( T \) diffusion steps, the signal ultimately deteriorates into isotropic Gaussian noise \( s_T \). To ensure the smoothness and gradual nature of this transformation, the signal \( s_t \) after \( t \) steps of forward diffusion follows the conditional distribution:
\begin{equation}
q\left( {{s_t}|{s_{t - 1}}} \right) = {\cal N}({v_t}{s_{t - 1}},(1 - v_t^2){\rm{I}})
\end{equation}
where \(\nu_t \propto {1/t}\) is the diffusion coefficient, and \( I \) represents the identity matrix. As the diffusion steps progress, \(\nu_t\) gradually decreases, diminishing the dominance of the signal. Consequently, \( s_t \) becomes increasingly noisy until, at the final step \( T \), it is fully transformed into Gaussian noise. Due to the Markov property, \( s_t \) can be derived from its initial state \( s_0 \):
\begin{equation}
q\left( {{s_t}|{s_0}} \right) = {\cal N}({\mu _t}{s_0},\sigma _t^2{\rm{I}})
\end{equation}
where \(\mu _t^{} = \Pi _{s = 1}^t{v_s}\) and \(\sigma _t^2 = 1 - \Pi _{s = 1}^tv_s^2\). Then, \( s_t \) can be represented as:
\begin{equation}
{s_t} = {\mu _t}{s_0} + {\sigma _t}\varepsilon
\label{eq5}
\end{equation}
where \(\varepsilon  \sim {\cal N}(0,I)\), which serves as the source of self-supervised information.

The reverse process serves as the inverse of the forward diffusion process, reconstructing the original signal \( s_0 \) from the noise distribution \( s_T \):
\begin{equation}
p\left( {{s_0}} \right) = \int {p({s_0}|{s_1})}  \cdots p({s_{T - 1}}|{s_T})d{s_1} \cdots d{s_T}
\end{equation}
Each step in the reverse process aims to approximate the true posterior distribution:
\begin{equation}
p\left( {{s_{t - 1}}|{s_t}} \right) \to q\left( {{s_{t - 1}}|{s_t},{s_0}} \right) = \frac{{q\left( {{s_t}|{s_{t - 1}}} \right)q\left( {{s_{t - 1}}|{s_0}} \right)}}{{q\left( {{s_t}|{s_0}} \right)}}
\end{equation}
By the Gaussian product theorem, we derive the following:
\begin{equation}
q\left( {{s_{t - 1}}|{s_t},{s_0}} \right) = {\cal N}({\mu _q}({s_t},{s_0}),{\sigma _q}{\rm{I}})
\end{equation}
where, 
\begin{equation}
\fontsize{9pt}{12pt}\selectfont 
{\mu _q}({s_t},{s_0}) = \frac{{\left( {1 - v_t^2} \right) \cdot {{\left( {\prod _{s = 1}^{t - 1}{v_s}} \right)}^2}}}{{1 - \prod _{s = 1}^{t - 1}v_s^2}}{s_0} + \frac{{{v_t}\left( {1 - \prod _{s = 1}^{t - 1}v_s^2} \right)}}{{1 - \prod _{s = 1}^tv_s^2}}{s_t}
\end{equation}
\begin{equation}
{\sigma _{q = }}\frac{{1 - \prod _{s = 1}^{t - 1}v_s^2}}{{1 - \prod _{s = 1}^tv_s^2}}\left( {1 - v_t^2} \right)
\end{equation}

The key to the reverse denoising process lies in predicting the mean of the posterior distribution. According to eq. (\ref{eq5}), the predicted original signal can be expressed as:
\begin{equation}
{\hat s_0} = \frac{1}{{\prod _{s = 1}^t{v_s}}}({s_t} - \sqrt {1 - \prod _{s = 1}^t(v_s^2)} {\delta _\theta }({s_t},t))
\end{equation}

Based on the self-supervised information from the forward process, the loss function of MSDGM can be expressed as follows:
\begin{equation}
{{\mathcal L}_{\rm{M}}} = \mathop {\rm{E}}\limits_{{s_0},\varepsilon ,t} {\left\| {\varepsilon  - {\delta _\theta }({s_t},t)} \right\|^2}
\end{equation}

The step-wise process in unsupervised training stage captures the structural semantics of modulated signals. The rich, scale-consistent features that are robust to variations in signal length and distribution, thereby supporting effective generalization.
\subsection{Diffusion-aware Feature Fusion Module}
In MSDGM, signals undergo progressive encoding and decoding across multiple hierarchical levels, inherently capturing features across a range of semantic and temporal scales. These features span from fine-grained local patterns to abstract global structures, providing a rich and diverse representation of the signal. However, directly utilizing features from a single layer often leads to suboptimal performance due to incomplete semantic coverage or scale bias. To address this, we propose DAFFus, which is designed to adaptively aggregate and align multi-scale features extracted from different layers of the diffusion network. By explicitly modeling the scale-aware nature of the diffusion process, DAFFus enhances the model’s ability to capture both local discriminative details and global structural information. This not only increases feature diversity and robustness, but also reinforces the representational capacity of the model in various communication scenarios.


As a classical encoder-decoder architecture, U-Net is employed as the noise prediction network in Fig. \ref{OverallStructure}, where its various layers effectively capture signal features ranging from fine-grained local details to high-level global representations. DAFFus takes \( s_t \) as input and extracts multi-layer features to enable hierarchical representation learning:
\begin{equation}
{{\cal F}_s} = \left\{ {f_s^{{b_i}} = {\cal A}{\cal P}\left( {{\cal U}_{\theta *}^{{{\rm{b}}_i}}\left( {{s_t}} \right)} \right)|i \in 1, \cdots ,L} \right\}
\end{equation}
where AP denotes the operation of pooling the final dimension to 1. \(\mathcal{U}_{\theta *}^{{{\text{b}}_{i}}}\) represents the trained MSDGM model, while \( b_i \) refers to the feature representation extracted at the \( i \)-th layer, spanning a total of \( L \) layers. This process effectively captures a ``semi-reconstructed" representation at the mid-stage, yielding a more compact semantic embedding. 

To maximize the potential of these multi-level features, integrating both low-level and high-level characteristics is a promising approach. 
The decoder path of the U-Net not only restores spatial resolution but also integrates semantic features from deeper layers with fine-grained details from shallower layers. We perform multi-scale feature concatenation along the channel dimension to effectively fuse information across different semantic levels:
\begin{equation}
{\cal F}_s^c = {\cal C}\left[ {f_s^{{b_{L/2+1}}}, \cdots ,f_s^{{b_L}}} \right]
\end{equation}
where \(\mathcal{C}[\cdot]\) denotes the concatenation operation along the channel dimension. Next, to make the information in \(\mathcal{F}_{s}^{c}\) more compact, we apply the following operation:
\begin{equation}
{\cal F}_s^D = \sigma \left( {W{\cal F}_s^c + b} \right)
\end{equation}
where \(\mathcal{F}_{s}^{D} \in \mathbb{R}^{d}\) represents the discriminative feature of signal \( s \), while \( W \) and \( b \) are trainable weights and biases, respectively. \( \sigma \) denotes the activation function. 
The recognition probability space is denote as:
\begin{equation}
p\left( {y = c|s} \right) = \frac{{\exp (W_{cls}^c{\cal F}_s^D + b_{cls}^c)}}{{\sum\nolimits_{c' = 1}^C {\exp (W_{cls}^{c'}{\cal F}_s^D + b_{cls}^{c'})} }}
\end{equation}
where \( W_{cls}^{c} \) and \( b_{cls}^{c} \) represent the classifier’s weight and bias corresponding to modulation type \( c \), respectively. The classification loss is:
\begin{equation}
{{\cal L}_{cls}} =  - {y_t}\log p(y = {y_t}|s)
\end{equation}

Finally, the loss is backpropagated to update both DAFFus and the classifier. At this point, DAFFus effectively integrates signal features from multiple hierarchical levels into a highly efficient and discriminative representation, providing accurate recognition for AMC.

\section{Experiments}

\begin{table*}[tbp]
\renewcommand{\arraystretch}{1.5}
\caption{Details of RML2016.10A, RML2016.10B, RML2018.01A and RML2022 datasets.
	\label{Details_dataset}}
\centering
\resizebox{17.5cm}{!}{
\begin{tabular}{cccccc}
\toprule
Dataset&	Signal format&	SNR& \makecell{number/SNR/type\\ in training set} &	\makecell{number/SNR/type\\ in test set}&	Modulation types\\
\midrule
RML2016.10A\cite{o2016radio}&
2×128&	-20 $\sim $18dB&	800&	200&\makecell{8PSK, AM-DSB, AM-SSB, BPSK, CPFSK, GFSK,\\ PAM4, QAM16, QAM64, QPSK, WBFM}\\
\midrule
RML2016.10B \cite{oshea_convolutional_2016}&
2×128&	-20$\sim $18dB&	4800&	1200&\makecell{8PSK, AM-DSB, BPSK, CPFSK, GFSK, PAM4,\\ QAM16, QAM64, QPSK, WBFM}\\
\midrule
RML2018.01A \cite{oshea_over--air_2018}&
2×1024&	-20$\sim $30dB&	3276&	819&\makecell{OOK, 4ASK, 8ASK, BPSK, QPSK, 8PSK, 16PSK, 32PSK, 16APSK, 32APSK,\\ 64APSK, 128APSK, 16QAM, 32QAM, 64QAM, 128QAM, 256QAM, AM-SSB-WC,\\ AM-SSB-SC, AM-DSB-WC, AM-DSB-SC, FM, GMSK, OQPSK}\\
\midrule
RML2022  \cite{sathyanarayanan_rml22_2023}&	2×128&	-20$\sim $20dB	&1600	&400&\makecell{	8PSK, AM-DSB, AM-SSB, BPSK, CPFSK, GFSK,\\ PAM4, QAM16, QAM64, QPSK, WBFM}\\
\bottomrule
\end{tabular}}
\end{table*}
\subsection{Experiment Settings}
\begin{figure*}[!t]
	\centering
	\captionsetup{skip=0pt}
	\setlength{\abovecaptionskip}{0cm}
	\setlength{\abovecaptionskip}{0cm}
	\includegraphics[width=1.0\textwidth]{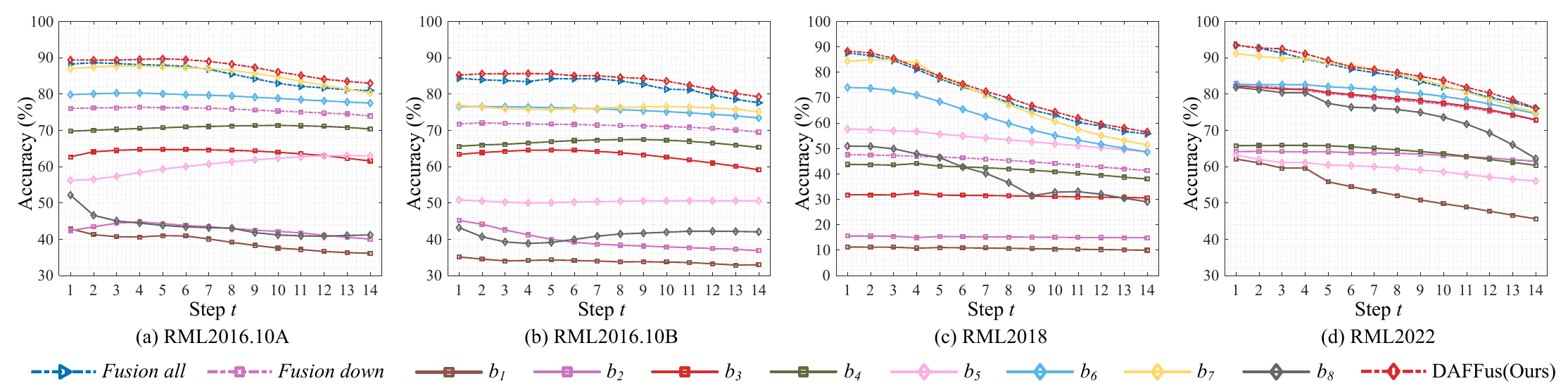}
	\caption{Ablation of different step $t$ and Block $b_{i}$ on RML2016.10A, RML2016.10B, RML2022, RML2018.01A with 10 labeled signals per type at 12dB.}
	\label{Abl_t_time}
		\vspace{-0.5cm}
\end{figure*}

To validate the effectiveness of ModFus-DM, we conduct experiments on four benchmark datasets: RML2016.10A \cite{o2016radio}, RML2016.10B \cite{oshea_convolutional_2016}, RML2018.01A \cite{oshea_over--air_2018}, and RML2022 \cite{sathyanarayanan_rml22_2023}, as summarized in Table \ref{Details_dataset}. RML2016.10A is the earliest and most widely adopted AMC dataset, while RML2016.10B extends it with more samples. RML2018.01A includes 24 modulation types, encompassing higher-order schemes such as 128APSK and 128QAM, and was generated under relatively benign environments. RML2022 refines RML2016.10A by addressing parameter inconsistencies and ad-hoc settings, offering more realistic data. 

For MSDGM, the total diffusion step $T$ is set to 100, trained over 2000 epochs with a learning rate of 0.0002 using the AdamW optimizer. For DAFFus, the feature dimension $d$ is set to 128, and the number of fusion layers $L$ is set to 8. DAFFus is trained for 50 epochs with Adam optimizer, and the learning rate decays from 0.01 to 0 via cosine annealing.

During the self-supervised modulation representation learning stage, the entire training set is treated as unlabeled to train the MSDGM. In the subsequent feature fusion stage, MSDGM was frozen, and DAFFus was updated using $N$ labeled signals per type per SNR. For fair evaluation, ten Monte Carlo experiments were conducted for $N=2,5,10,20$ settings, and the average accuracy was reported.
\subsection{Ablation Study}
\subsubsection{Ablation of DAFFus and $s_t$}
\begin{figure*}[!t]
	\centering
	\captionsetup{skip=0pt}
	\setlength{\abovecaptionskip}{0cm}
	\setlength{\abovecaptionskip}{0cm}
	\includegraphics[width=1.0\textwidth]{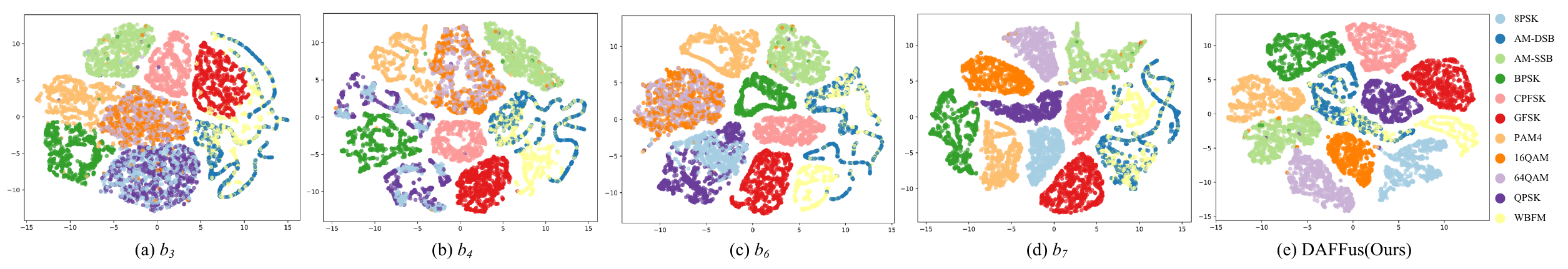}
	\caption{t-SNE of representations from different blocks $b_i$ and DAFFus when fixed $s_1$ as input on 12dB RML2016.10A dataset.}
	\label{tSNE_Blocks}
		\vspace{-0.5cm}
\end{figure*}

To validate the effectiveness DAFFus module within the ModFus-DM, we performed ablation studies on  diffusion feature selection and fusion. Results are presented in Fig. \ref{Abl_t_time}, with visualizations in Fig. \ref{tSNE_Blocks}. Experiments were conducted at SNR=12dB, where only 10 labeled signals per type to train DAFFus. Notably, MSDGM remained frozen during DAFFus training. Specifically, $b_i$ denotes features from U-Net block $b_i$. \textit{Fusion all} and \textit{fusion down} refer to fusing features from all blocks ($b_1$–$b_8$) and from the downsampling blocks ($b_1$–$b_4$), respectively. DAFFus fuses diffusion features from the upsampling blocks ($b_5$–$b_8$) to derive modulation representations.

As shown in Fig. \ref{Abl_t_time}, DAFFus achieves the best performance in over 96\% settings. For diffusion features from single block $b_i$, when using the signal $s_1$ generated at forward diffusion step $t=1$ as input, DAFFus outperforms the second-best $b_7$, by 2.4\%, 8.3\%, 3.9\%, and 2.3\% on the RML2016.10A, RML2016.10B, RML2018, and RML2022 datasets, respectively. Fig. \ref{tSNE_Blocks} further shows DAFFus yields the clearest clustering, confirming its superior feature discrimination. From a feature fusion perspective, DAFFus consistently surpasses both the \textit{Fusion down} and \textit{Fusion all}. With $s_1$ as input, DAFFus improves accuracy by 13.36\% and 1.12\% compared to \textit{Fusion down} and \textit{Fusion all}, respectively. The subpar performance of \textit{Fusion down} is due to information loss from downsampling, producing coarser features.\textit{Fusion all} suffers from low-quality downsampling features diluting overall representation quality, making it less effective than DAFFus.

The diffusion features extracted from higher-level blocks contain richer modulation representations. In most cases, the features from $b_6$ and $b_7$ yield the highest recognition accuracy among all single block features. This is due to skip connections enabling integration of both upsampling and downsampling information, producing more comprehensive and discriminative representations. In contrast, features from $b_8$ are more tailored to generative tasks, focusing primarily on reconstruction fidelity, which in turn weakens their discriminative capacity.

\begin{table}[tbp]
	\renewcommand{\arraystretch}{1.5}
	\caption{Ablation study on the total diffusion steps $T$ is conducted on RML2016.10A. "Mean" denotes the average accuracy across various SNR.
		\label{Abl_T}}
	\centering
	\resizebox{7cm}{!}{
		\begin{tabular}{c|cccccc}
			\toprule
			$T$	&0dB&	4dB&	8dB&	12dB&	16dB&	Mean\\
			\midrule
			10&	74.68&	79.97&	81.68&	82.62&	81.27&	80.04\\
			30&	80.14&	82.14&	82.51&	82.28&	80.98&	81.61\\
			50&	84.18&	83.88&	83.38&	83.75&	82.65&	83.57\\
			70&	87.45&	87.17&	88.75&	87.23&	86.27&87.37\\
			100&	90.65&	92.93&	93.26&	93.96&	92.48&	89.29\\
			150&89.45& 91.03& 90.76& 89.95& 88.33& 89.90\\
			200&89.40&91.14&91.19&89.82&87.99&89.91\\
			\bottomrule
	\end{tabular}}
	\vspace{-0.5cm}
\end{table}

Low-noise input signals yield better recognition. For diffusion features extracted using the same method, as the diffusion step $t$ increases, noise dominates and structural information fades, causing accuracy to drop. Signal $s_t$ obtained at smaller diffusion steps preservers more of the original structural characteristics, providing richer modulation information. Moreover, recognition performance remains nearly consistent across small $t$ values. For example, on RML2016.10A in Fig. \ref{Abl_t_time}(a), when $t$ range from 1 to 7, the recognition accuracy fluctuates by only 0.7\%. Therefore, we fix $t = 1$ for all subsequent experiments.

DAFFus is both necessary and effective. On the one hand, ingle-block diffusion features vary in discriminative power, making it hard to pick the best one. This challenge is especially evident on the RML2016.10B dataset in Fig. \ref{Abl_t_time}(b), where the diffusion features from $b_6$ and $b_7$ alternately yield the best single block recognition results. On the other hand, by fusing multi-scale diffusion features from different blocks, DAFFus effectively overcomes the difficulty of manually selecting the optimal single block diffusion feature and significantly improves model performance.

\subsubsection{Ablation of total diffusion steps $T$} 
To evaluate the impact of total diffusion steps $T$ on modulation representation quality, we conducted ablation studies with $T$ ranging from 10 to 200. MSDGM was trained on the standard RML2016.10A dataset. Subsequently, for the frozen MSDGM, 10 labeled signals per class per SNR were used to train DAFFus. The final recognition results are presented in Table \ref{Abl_T}.

Recognition performance of ModFus-DM improves as total diffusion steps $T$ increase. Specifically, when $T$ increases from 10 to 100, the recognition accuracy steadily improves. At $T=100$, the average accuracy surpasses that of $T=10$ by 9.86\%, indicating a substantial enhancement in representation quality. However, when $T$ exceeds 100, the performance gains become marginal. For instance, increasing $T$ from 100 to 200 results in only a 0.62\% improvement, suggesting that the representation quality has reached saturation. These findings indicate that a larger $T$ facilitates the modeling of more intricate structural characteristics in the signals, thereby boosting the discriminative power and semantic richness of the learned diffusion features. Nonetheless, an excessively large $T$ does not significantly enhance discriminative information and may instead introduce unnecessary computational overhead. Balancing performance and efficiency, we adopt $T=100$ in all subsequent experiments.

\begin{figure*}[!t]
	\centering
	\captionsetup{skip=0pt}
	\setlength{\abovecaptionskip}{0cm}
	\setlength{\abovecaptionskip}{0cm}
	\vspace{-0.2cm}
	\includegraphics[width=0.95\textwidth]{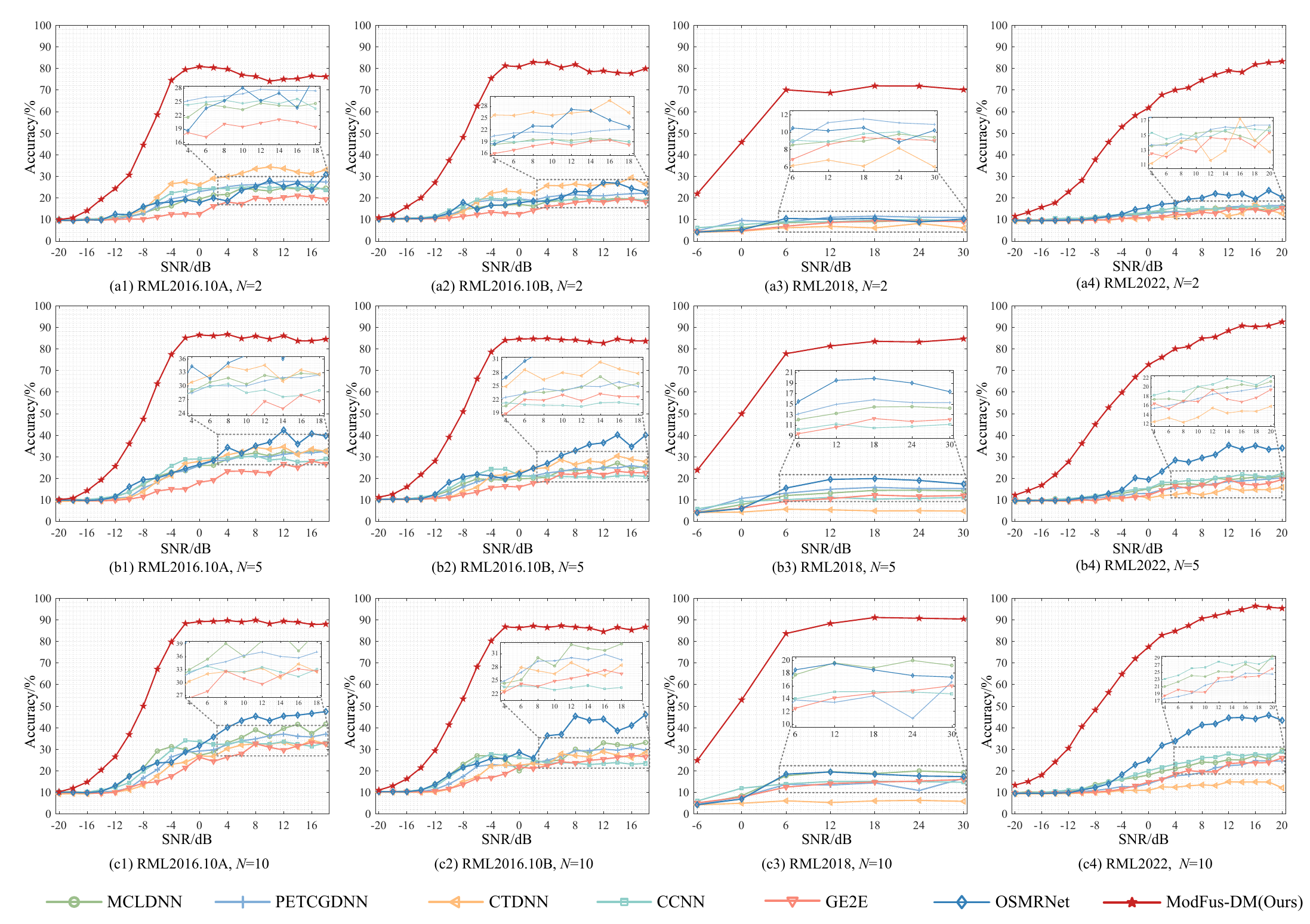}
	\caption{Comparision with supervised AMC methods at $N=2, 5, 10$ on RML2016.10A, RML2016.10B, RML2018 and RML2022.}
		\vspace{-0.5cm}
	\label{Com_supervised}
\end{figure*}

\subsection{Comparison with Supervised AMC Methods}
To assess the performance of the proposed ModFus-DM under limited labeled signal scenarios, we compared it with several supervised AMC models, including MCLDNN \cite{xu_spatiotemporal_2020}, CTDNN \cite{kong_transformer-based_2021}, PETCGDNN \cite{zhang_efficient_2021}, GE2E \cite{zhang_open_2022}, CCNN \cite{liu_automatic_2023}, and OSMRNet \cite{ling_osmr_2024}. Fig. \ref{Com_supervised} presents the recognition performance of ModFus-DM and the supervised methods on RML2016.10A, RML2016.10B, RML2018.01A, and RML2022, using only 2, 5, or 10 labeled signals per class. For reference, Table \ref{Com_sup} provides the performance of supervised models trained on fully labeled datasets. Notably, ModFus-DM achieves competitive performance with as few as 10 or 20 labeled signals per class.


\begin{table*}[tbp]
	\renewcommand{\arraystretch}{1.5}
	\caption{Comparison with supervised AMC methods at 0dB and 12dB. Notably, all supervised models are trained on all labeled signals in the corresponding dataset. However, ModFus-DM only uses 10, 20 and all labeled signals per type. $\uparrow$ means the accuracy of ModFus-DM (all) higher than that of the compared method.
		\label{Com_sup}}
	\centering
	\resizebox{18cm}{!}{
\begin{tabular}{c|cc|cc|cc|cc}
\toprule
\multirow{2}{*}{Methods} & \multicolumn{2}{c|}{RML2016.10A} & \multicolumn{2}{c|}{RML2016.10B} & \multicolumn{2}{c|}{RML2018.01A} & \multicolumn{2}{c}{RML2022} \\
& 0dB & 12dB & 0dB & 12dB & 0dB & 12dB & 0dB & 12dB \\
\midrule
PETCGDNN(ALL)\cite{zhang_efficient_2021} &78.18(15.09$\uparrow$)&	86.14(9.91$\uparrow$)&	87.88(5.19$\uparrow$)&	91.68(2.07$\uparrow$)	&38.64(22.67$\uparrow$)&	70.41(25.43$\uparrow$)&	77.59(8.98$\uparrow$)&	90.11(8.89$\uparrow$)	\\
CTDNN(ALL)\cite{kong_transformer-based_2021} &90.91(2.36$\uparrow$)&	92.09(3.96$\uparrow$)&	91.76(1.31$\uparrow$)&	93.44(0.31$\uparrow$)&57.47(3.84$\uparrow$)&91.68(4.16$\uparrow$)&	83.34(3.23$\uparrow$)&	96.64(2.36$\uparrow$)\\
CCNN(ALL)\cite{liu_automatic_2023} &68.05(25.22$\uparrow$)&	76.18(19.87$\uparrow$)&	69.20(23.87$\uparrow$)&	79.12(14.63$\uparrow$)&30.91(30.40$\uparrow$) &		44.05(51.79$\uparrow$)&	66.98(19.59$\uparrow$)&	84.07(14.93$\uparrow$)\\
GE2E(ALL)\cite{zhang_open_2022} &88.91(4.36$\uparrow$)&	92.64(3.41$\uparrow$)&	91.07(2.00$\uparrow$)&	93.64(0.11$\uparrow$)	&49.03(12.28$\uparrow$)&	88.21(7.63$\uparrow$)&	81.93(4.64$\uparrow$)&	96.30(2.70$\uparrow$)\\
OSMRNet(ALLl)\cite{ling_osmr_2024} & 67.55(25.72$\uparrow$)&	84.45(11.60$\uparrow$)&	78.96(14.11$\uparrow$)&	92.24(1.51$\uparrow$)	&33.04(28.27$\uparrow$) &	84.07(11.77$\uparrow$)&	71.14(15.43$\uparrow$)&	87.98(11.02$\uparrow$)\\
\midrule
ModFus-DM($N=10$)&89.09(4.18$\uparrow$)&	89.34(6.71$\uparrow$)&	86.30(6.77$\uparrow$)&	84.47(9.28$\uparrow$)&	52.84(8.47$\uparrow$)&	88.27(7.57$\uparrow$)&	77.39(9.18$\uparrow$)&	93.42(5.58$\uparrow$)\\
ModFus-DM($N=20$)&91.52(1.75$\uparrow$)&	91.08(4.97$\uparrow$)&	89.60(3.47$\uparrow$)&	87.44(6.31$\uparrow$)&	55.62(5.69$\uparrow$)&	92.07(3.77$\uparrow$)&	80.58(5.99$\uparrow$)&	97.09(1.91$\uparrow$)\\
\midrule
ModFus-DM(ALL)&\textbf{93.27}&	\textbf{96.05}&	\textbf{93.07}&	93.75&	\textbf{61.31}&	\textbf{95.84}&	\textbf{86.57}&	\textbf{99.00}\\
\bottomrule
\end{tabular}}
\vspace{-0.5cm}
\end{table*}

\subsubsection{Comparison with supervised AMC models using limited labeled signals} As illustrated in Fig. \ref{Com_supervised}, under limited labeled scenarios ($N$=2,5,10), the proposed ModFus-DM significantly outperforms traditional supervised AMC methods. Specifically, on the RML2016.10A dataset across various SNRs (0–18dB), ModFus-DM surpasses the second-best method, CTDNN, by 39.68\%–54.84\% at $N$=2 (a1), 50.15\%–58.63\% at $N$=5 (b1), and 53.52\%–62.63\% at $N$=10 (c1). The performance advantage becomes even more pronounced on the more challenging RML2018.01A dataset. At 18dB, ModFus-DM achieves recognition accuracies of 71.88\%, 83.46\%, and 91.01\% for $N$=2,5,10, respectively. In contrast, all supervised methods exhibit substantial degradation, with accuracies ranging merely from 6.06\% to 19.89\%. This decline can be attributed to the increased complexity of RML2018.01A, which contains 24 modulation types and a signal length of 1024 (twice that of the other three datasets). The richer modulation patterns and extended temporal structure substantially increase learning difficulty for supervised models, resulting in sharp drops in recognition accuracy. However, ModFus-DM benefits from longer signal sequences, which enhances its ability to capture temporal and structural information during the self-supervised modulation representation learning stage, thereby ensuring robust feature extraction and maintaining high recognition accuracy with scarce labels.

The recognition accuracy of both ModFus-DM and supervised AMC models consistently improves as the number of labeled signals $N$ increases. For instance, on the RML2016.10A dataset at 12dB, MCLDNN achieves an accuracy of 39.66\% at $N$=10, representing a 16.46\% improvement over its performance at $N$=2. For supervised AMC models, this performance gain primarily stems from the direct supervision provided by label information. As the number of labeled signals increases, the models are better able to learn the statistical distributions of different modulation types. On the RML2016.10B dataset at 12dB, ModFus-DM attains an accuracy of 84.48\% at $N$=10, which is 5.63\% higher than that at $N$=2. The availability of more labeled signals enables DAFFus to more effectively learn how to fuse modulation characteristics from the semantically rich diffusion features, thereby achieving more accurate recognition.

\subsubsection{Comparison with fully-supervised AMC models} As shown in Table \ref{Com_sup}, ModFus-DM demonstrates competitive, and in some cases superior performance compared to supervised AMC methods, with only 10 or 20 labeled signals. Moreover, when ModFus-DM is fine-tuned using the full set of labeled data, it nearly surpasses all supervised models. On the RML2016.10A at 0dB, ModFus-DM($N$=10) achieves an accuracy of 89.09\%, which is only 1.82\% lower than the best-performing CTDNN(ALL) (90.91\%), while utilizing merely 1.25\% of the labeled data. On RML2016.10B at 12dB, ModFus-DM($N$=20) achieves 87.44\% accuracy, surpassing several supervised methods (e.g., outperforming CCNN(ALL) by 8.32\%) and approaching the performance of MCLDNN(ALL), despite using only 0.42\% of the labeled signals. Benefiting from the longer signal in the RML2018.01A dataset, ModFus-DM($N$=20) reaches 92.07\% accuracy, trailing MCLDNN(ALL) by 3.66\% while outperforming all other methods. Moreover, when DAFFus trained with all available labeled samples, ModFus-DM(ALL) surpasses all supervised AMC methods. On the synthetic RML2022 at 12dB, where channel conditions are relatively ideal, ModFus-DM($N$=20) surpasses all supervised methods by margins ranging from 0.45\% to 9.11\%.

These results confirm that ModFus-DM maintains robust performance under limited label scenarios across diverse signal lengths and benchmark datasets, highlighting its strong generalizability and practical applicability in complex and dynamic real-world communication environments.

\subsection{Comparison with Self- and Semi-supervised AMC Methods}
\begin{figure*}[!t]
	\centering
	\captionsetup{skip=0pt}
	\setlength{\abovecaptionskip}{0cm}
	\setlength{\abovecaptionskip}{0cm}
	\includegraphics[width=0.95\textwidth]{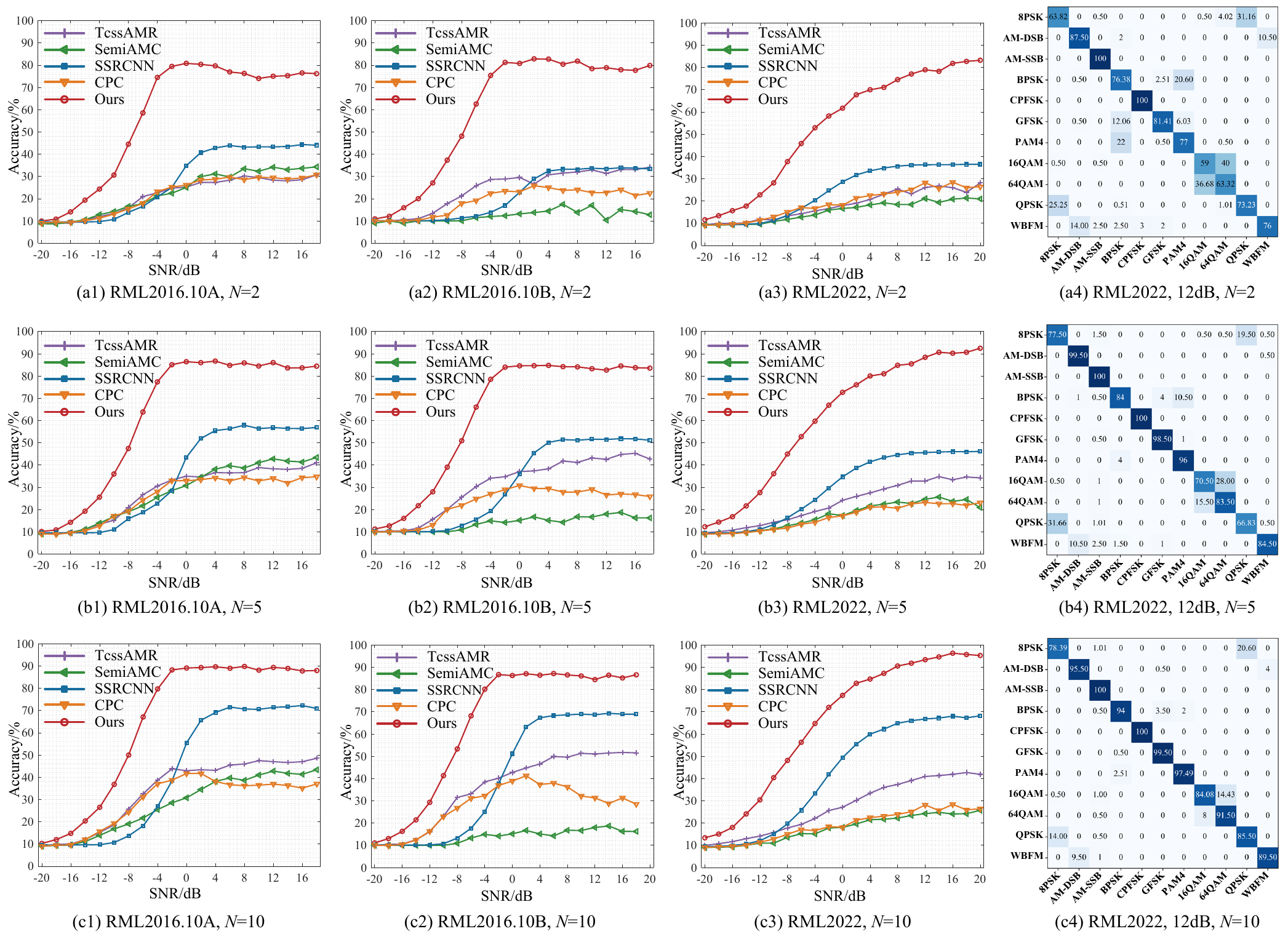}
	\caption{Comparision with self- and semi-supervised AMC methods at $N$=2, 5, 10 on RML2016.10A, RML2016.10B and RML2022.}
	\label{Com_semi}
	\vspace{-0.5cm}
\end{figure*}

To evaluate the performance of ModFus-DM under limited label scenarios, we conducted comparative experiments against several representative self- and semi-supervised AMC methods, including TcssAMR \cite{kong_transformer-based_2023}, SemiAMC \cite{liu_self-contrastive_2021}, SSRCNN \cite{dong_ssrcnn_2021} and CPC \cite{CPC_2018}. Specifically, when only 2, 5, or 10 labeled signals per type are available, the recognition results across various SNRs on the RML2016.10A, RML2016.10B, and RML2022 datasets are presented in Fig. \ref{Com_semi}.

The proposed ModFus-DM consistently outperforms mainstream self- and semi-supervised methods across varying SNRs and different numbers of labeled signals. Under extremely limited supervision with $N$=2, ModFus-DM surpasses the second-best method, SSRCNN, by 33.16\%, 49.60\%, and 38.93\% at 8dB on the RML2016.10A, RML2016.10B, and RML2022 datasets, respectively, demonstrating its strong limited label performance. As the number of labeled signals per type $N$ increases, recognition accuracy improves across all methods, with ModFus-DM showing particularly substantial gains. In high SNR scenarios (SNR$\ge$8dB), ModFus-DM achieves recognition accuracies exceeding 88.14\%, 84.48\%, and 90.57\% on the RML2016.10A, RML2016.10B, and RML2022 datasets, respectively. As shown in the confusion matrix of Fig. \ref{Com_semi}(c4), when using only 10 labeled signals per type which is just 0.625\% of the full training set, ModFus-DM attains over 90\% accuracy for most modulation types. Although 8PSK, 16QAM, 64QAM, and QPSK exhibit minor confusion due to similar modulation mechanisms, limited labeled signals, and constrained signal lengths, their recognition accuracies still remain above 78.39\%, further underscoring the model’s strong capability to distinguish complex modulation structures.

It is noteworthy that SSRCNN consistently achieves the second-best performance in most scenarios, primarily due to its single-stage semi-supervised training paradigm. Both labeled and unlabeled signals are jointly fed into the model for optimization, enabling the network to leverage label information early in training to guide the formation of discriminative decision boundaries. Consequently, it performs relatively well under limited-label conditions. However, this paradigm makes the model highly dependent on the quality of labeled signals and often limits its generalization capability. In contrast, methods such as SemiAMC, CPC, and TcssAMR follow a two-stage training framework. They first pre-train a feature extractor on unlabeled data, followed by fine-tuning both the feature extractor and the classifier using labeled signals. However, the absence of task-specific supervision during pre-training poses a significant challenge, making it difficult for the model to learn discriminative features from complex modulated signals. Moreover, the limited number of labeled signals in the fine-tuning stage is insufficient to compensate for this deficiency, ultimately constraining their performance.

The core distinction of ModFus-DM lies in the fine-tuning phase, during which only the classifier parameters are updated while the feature extractor MSDGM remains frozen. This design effectively mitigates the risk of overfitting caused by the scarcity of labeled signals. More importantly, during the pre-training phase, the model reconstructs the modulation structure and distributional characteristics of the signals step by step, thereby uncovering rich modulation semantics. This label-free modulation representation learning approach enables ModFus-DM to achieve superior recognition performance with a minimal amount of labeled signals.

\subsection{Generalization Performance}
\begin{figure*}[!t]
	\centering
	\captionsetup{skip=0pt}
	\setlength{\abovecaptionskip}{0cm}
	\setlength{\abovecaptionskip}{0cm}
	\includegraphics[width=0.95\textwidth]{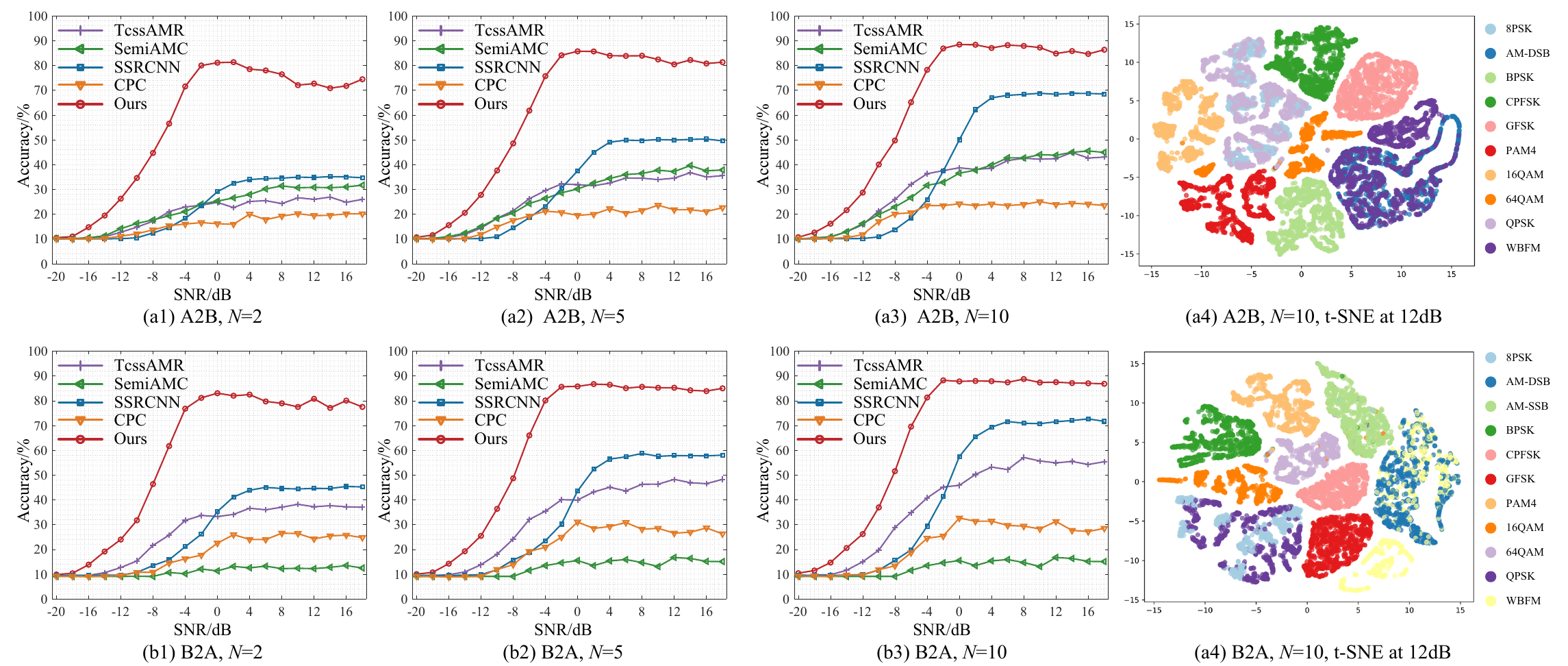}
	\caption{Generalization performance comparison when training and testing datasets are different. (a1)–(a3) show results when trained on RML2016.10A and tested on RML2016.10B (A2B) with different numbers of labeled samples per class ($N$=2, 5, 10). (b1)-(b3) correspond to the reverse setup (B2A). (a4) and (b4) illustrate the t-SNE visualizations of the features by ModFus-DM under A2B and B2A settings at 12dB, respectively.}
	\label{Com_gen_semi}
	\vspace{-0.5cm}
\end{figure*}

To ModFus-DM’s generalization under distribution shift, we compare it with mainstream self- and semi-supervised methods: TcssAMR \cite{kong_transformer-based_2023}, SemiAMC \cite{liu_self-contrastive_2021}, SSRCNN \cite{dong_ssrcnn_2021} and CPC \cite{CPC_2018}. Results are shown in Fig. \ref{Com_gen_semi}. Specifically, Fig. \ref{Com_gen_semi}(a)–(c) present the performance of models trained on the RML2016.10A and tested on RML2016.10B (A2B), while Fig. \ref{Com_gen_semi}(d)–(e) show the reverse setting, where the models are trained on RML2016.10B and tested on RML2016.10A (B2A).

ModFus-DM consistently outperforms existing methods across various distribution shift scenarios and SNR levels. Notably, in both A2B and B2A settings, when SNR$>$0dB, ModFus-DM achieves recognition accuracies exceeding 70.84\%, while all other methods remain below 45.38\%, highlighting the significant advantage of ModFus-DM in cross-distribution recognition with limited labeled signals. As the number of labeled signals per class $N$ increases, ModFus-DM more effectively captures the discriminative modulation features, further enhancing recognition performance. At $N$=10, ModFus-DM achieves accuracies of 84.67\% and 86.81\% under A2B and B2A settings with SNR$>$0dB, respectively. In contrast, methods such as TcssAMR, SemiAMC, SSRCNN, and CPC demonstrate limited generalization, attributed to the poor representational capacity of their pretraining stages. Even under high-SNR and relatively sufficient supervision ($N$=10), their accuracy generally remains below 55\%, revealing inherent performance bottlenecks.  It is also worth noting that the single-stage training strategy of SSRCNN, the second-best method, results in strong sensitivity to label quantity. For instance, in the A2B setting at SNR=10dB, SSRCNN's accuracy improves markedly from 34.96\% at $N$=2 to 68.77\% at $N$=10. Although the gain is notable, it comes at the cost of high annotation demand, limiting the method’s robustness in limited label scenarios.

\begin{figure}[!t]
	\centering
	\captionsetup{skip=0pt}
	\setlength{\abovecaptionskip}{0cm}
	\setlength{\abovecaptionskip}{0cm}
	\includegraphics[width=0.47\textwidth]{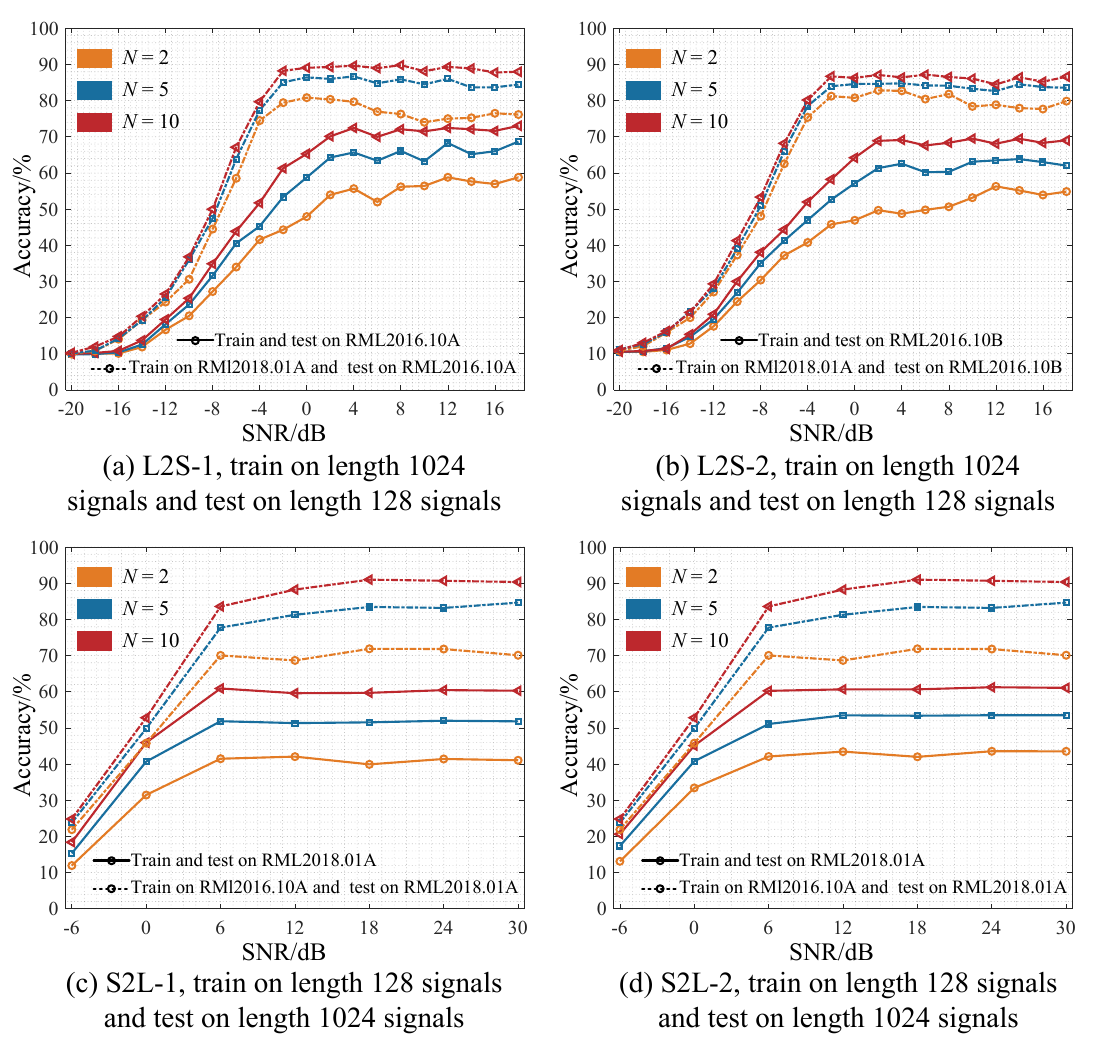}
	\caption{Performance of ModFus-DM under variable-length signal and different datasets scenarios. (a) and (b) correspond to long-to-short (L2S) settings, where the model is trained on 1024-length signals and tested on 128-length signals. (c) and (d) correspond to short-to-long (S2L) settings, where the model is trained on 128-length signals and evaluated on 1024-length signals. Three limited label conditions are considered: $N$=2,5,10.}
	\label{Temporal_flexibility}
	\vspace{-0.5cm}
\end{figure}

The remarkable generalization capability of ModFus-DM stems from the unique feature extraction mechanism. Leveraging self-supervised learning, it captures multi-scale, context-rich representations with strong structural modeling capabilities, making it inherently well-suited for abstracting complex modulation patterns and enabling cross-distribution generalization. The deep, multi-granular feature space it constructs offers robust resistance to distributional discrepancies between the source and target distributions.

\vspace{-0.15cm}
\subsection{Temporal Flexibility}

To evaluate ModFus-DM’s adaptability to variable-length signals, we conducted experiments using different signal lengths.  In Table \ref{Table_2018_varylenght}, the model is trained on 1024-length signals and tested on signals ranging from 64 to 1024, all randomly cropped from the RML2018.01A dataset. Fig. \ref{Temporal_flexibility} further explores a more challenging scenario where training and testing signals differ in both length and distribution.

\begin{table}[tbp]
	\renewcommand{\arraystretch}{1.5}
	\caption{Performance under variable-length signals for training and testing on 24dB RML2018.01A. The ModFus-DM was pre-trained on signals of length 1024 and evaluated on signals of lengths 64, 128, 256, 512, 768, and 1024. Notably, the signals of lengths 64 to 768 are randomly cropped from the full-length 1024 signals.
		\label{Table_2018_varylenght}}
	\centering
		\resizebox{8cm}{!}{
		\begin{tabular}{c|cccccc}
			\toprule
			Signal length &64&128&256&512&768&1024\\
			\midrule
			$N$=2 &35.53&50.48&65.23&70.96&	72.51&	71.82\\
			$N$=5 &40.92&62.03&76.40&84.37&83.23 &	83.19\\
			$N$=10 &45.83&	68.36&	83.90&	90.10&	91.09&	90.69\\
			\bottomrule
		\end{tabular}}
		\vspace{-0.5cm}
	\end{table}

As shown in Table \ref{Table_2018_varylenght}, recognition performance increases with signal length. Longer signals encompass more complete modulation patterns and richer semantic structures, which substantially enhance the model's recognition capability. For instance, under the $N$=10 setting, the recognition accuracy is only 45.83\% when the signal length is 64, primarily due to insufficient semantic information and limited labeled signals. As the length increases to 512, the signal contains enough semantic information, leading to a substantial accuracy improvement to 90.10\%. Beyond 512, the semantic information becomes sufficient for reliable recognition, and performance begins to saturate. Under both $N$=5 and $N$=10, the recognition accuracy across lengths of 512, 768, and 1024 varies only slightly, by 1.18\% and 0.99\%, respectively.

Furthermore, as illustrated in Fig. \ref{Temporal_flexibility}, ModFus-DM demonstrates strong adaptability even under the more complex scenario of cross-length and distribution shifts. In the train on long signals and test on short signals (L2S) setting ((a) and (b)), the recognition performance on short signal approaches its upper bound, defined as the accuracy achieved by the model trained and tested on short signals. For example, under the conditions of $N$=10 and 10dB, the model achieves 71.49\% and 69.51\% accuracy on RML2016.10A and RML2016.10B, respectively, which are only 16.65\% and 16.60\% lower than their distribution-matched upper bounds. By contrast, the train on short signals and test on long signals (S2L) setting ((c) and (d)) poses a greater challenge, as the model must generalize from short training signals to long signals containing richer structural information. Under the $N$=10 and 24dB, when trained on RML2016.10A and RML2016.10B, the model achieves 60.48\% and 61.26\% accuracy on RML2018.01A, respectively. This performance degradation is primarily due to short signals limit the model in capturing long-range dependencies, and, the constrained length of training signals results in decision boundaries based only on local features, which hampers generalization to the more complex global representations in longer signals.

\begin{figure}[!t]
	\vspace{-0.4cm}
	\centering
	\captionsetup{skip=0pt}
	\setlength{\abovecaptionskip}{0cm}
	\setlength{\abovecaptionskip}{0cm}
	\includegraphics[width=0.47\textwidth]{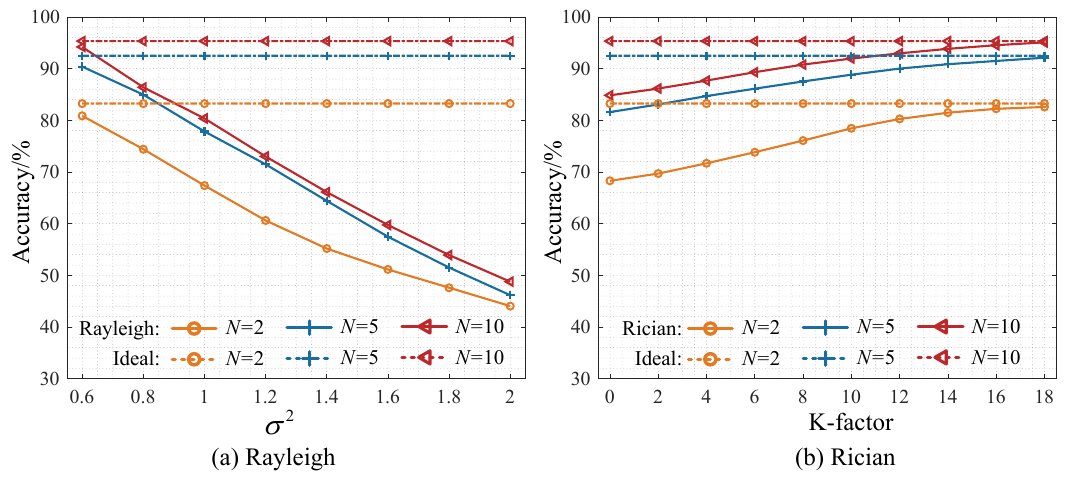}
	\caption{Performance under different channels. (a) Effect of ${\sigma ^2}$ on recognition accuracy under Rayleigh channel. (b) Effect of K-factor on recognition accuracy under Rician channel. “Ideal” mean the performance on signals without fading.}
	\label{Effect_channel}
	\vspace{-0.5cm}
\end{figure}

Overall, ModFus-DM demonstrates exceptional adaptability and robustness in handling variable-length signals. Whether transferring from long to short signals, generalizing from short to long, or operating under scenarios with discrepancies in both datasets and signal lengths, the model consistently delivers strong performance. This resilience is attributed to its multi-step diffusion modeling mechanism, which enables consistent modulation representation across signal durations, and to DAFFus dynamic fusion of multi-scale diffusion features. Together, they construct a high-quality modulation representation space that is both length-invariant and task-adaptive. ModFus-DM thus offers an effective and practical solution for non-fixed-length signal recognition in real-world communication systems.

\vspace{-0.15cm}
\subsection{Real-world Performance}

\begin{figure}[!t]
	\vspace{-0.4cm}
	\centering
	\captionsetup{skip=0pt}
	\setlength{\abovecaptionskip}{0cm}
	\setlength{\abovecaptionskip}{0cm}
	\includegraphics[width=0.47\textwidth]{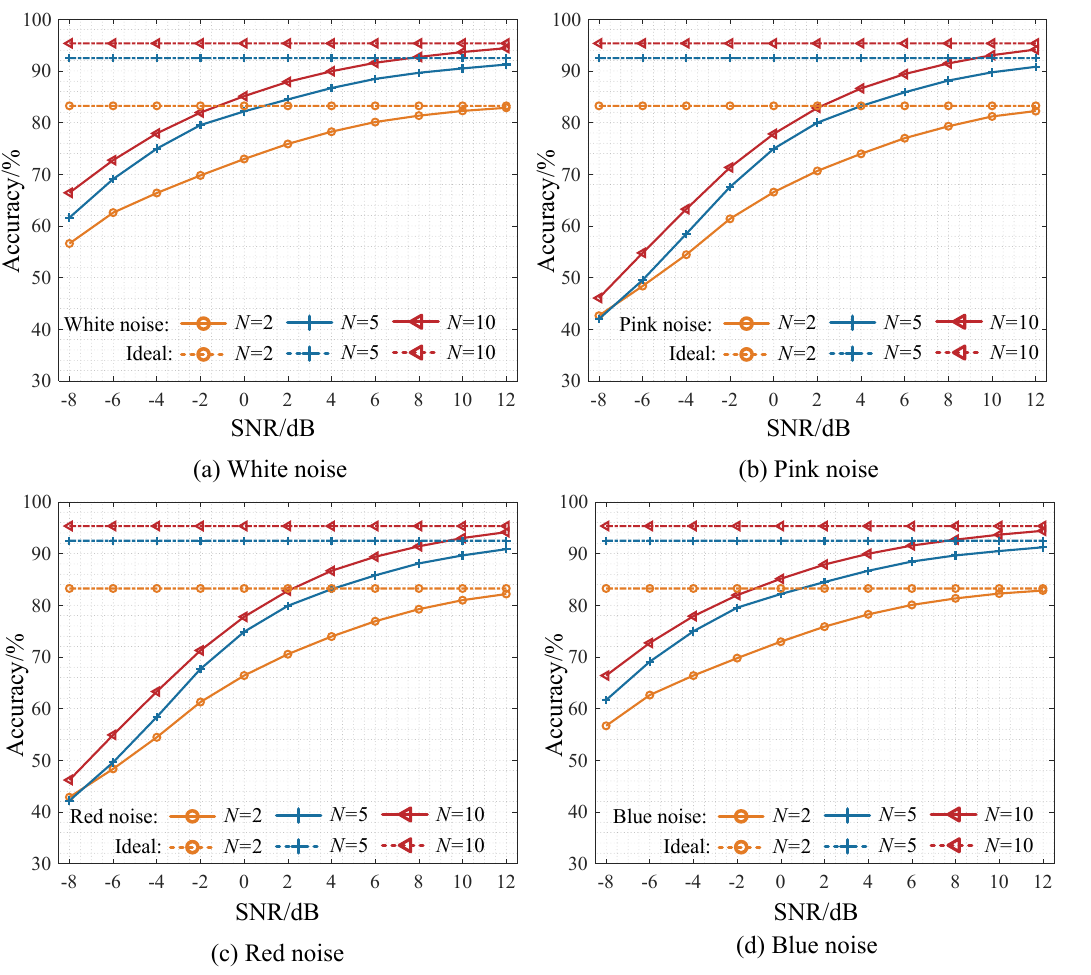}
	\caption{Performance under different colored noise. Effect of (a)White noise, (b)Pink noise, (c)Red noise, and (d)Blue noise on recognition accuracy. “Ideal” mean the performance on signals without extra noise.}
	\label{RealColored}
	\vspace{-0.5cm}
\end{figure}

To evaluate the performance of ModFus-DM under complex real-world wireless communication conditions, such as multipath interference and variations in line-of-sight propagation, we conducted experiments under Rayleighigh and Rician fading channels, as shown in Fig. \ref{Effect_channel}. To assess the model’s robustness to spectral distribution shifts and non-stationary interference, we tested it under various types of colored noise, with the results presented in Fig. \ref{RealColored}. In these evaluations, 20dB signals from RML2022 dataset were used as ideal reference signals.

\subsubsection{Performance under different channels fading} As shown in Fig. \ref{Effect_channel}(a), under Rayleighigh fading with $\sigma^2$ = 0.6 $\sim 1.2$, ModFus-DM maintains over 73\% accuracy using only 10 labeled signals. Notably, when $\sigma^2$ = 0.6,  its performance closely approaches that on clean signals, with accuracy gaps of just 2.37\%, 2.10\%, and 1.13\% for $N$ = 2, 5, 10, respectively. As $\sigma^2$ increases, fading distortion intensifies, leading to a gradual accuracy decline. Similarly, in Fig. \ref{Effect_channel}(b), when the Rician factor $K$ ranges from 2 to 18, ModFus-DM achieves over 81.58\% accuracy using just 5 labeled signals. Performance improves with increasing $K$, as stronger line-of-sight components enhance channel quality.

\subsubsection{Performance under various colored noise types} 
As illustrated in Fig. \ref{RealColored}, 
ModFus-DM demonstrates strong limited-label performance across various SNRs and colored noise types, even with limited labels. With just 2 labeled signals per type, ModFus-DM achieves over 66.38\% accuracy under white, pink, red, and blue noise. As the number of labeled signals increases to $N$=10, the performance further improves, with the average accuracy surpassing 77.75\% when SNR$>$0dB. Notably, Similar results under white and blue noise (Fig. \ref{RealColored}(a) and (d)) stem from their comparable low-frequency energy. Also, Fig. \ref{RealColored}(b) and (c) show that pink and red noise, which is characterized by significantly amplified low-frequency components. This make them introduce more severe interference to the predominantly low-frequency signals, thereby degrading recognition performance. 

Overall, by leveraging its diffusion-based reconstruction process in the self-supervised stage, ModFus-DM effectively learns robust semantic representations of modulated signals. This capability enhances its resilience under wireless fading channels and ensures adaptability and generalization in non-ideal noise environments. These strengths establish a solid foundation for the practical deployment of ModFus-DM in complex electromagnetic scenarios.
\vspace{-0.2cm}
\section{Conclusion}

We propose ModFus-DM, the first foundational model for AMC based on generative models. By employing an unsupervised generative learning paradigm, ModFus-DM effectively uncovers intrinsic structures and high-level semantic features in modulated signals. Its progressive denoising process captures multi-scale, temporal-spanning representations, forming robust feature spaces with strong discriminative power and generalization ability. Evaluations across variable signal lengths, limited labels, distribution shifts, and diverse channel conditions demonstrate that ModFus-DM consistently outperforms existing approaches on four benchmark datasets. These results confirm its superiority in feature quality, classification accuracy, and robustness. Overall, ModFus-DM offers a promising direction for the application of unsupervised generative models in wireless signal understanding and intelligent spectrum management.


However, ModFus-DM currently focuses on single-modality IQ signals, without fully exploring other informative modalities, such as spectrum and constellation diagrams. In future work, we will explore multi-modal modulation representation learning to further enhance the adaptability and scalability of ModFus-DM.
\vspace{-0.15cm}
\bibliography{IEEEabrv, ref} 	
\bibliographystyle{IEEEtran}
\end{document}